\documentclass[12pt]{article}

\usepackage[preprint,nonatbib]{nips_gibs}

\usepackage[utf8]{inputenc} 
\usepackage[T1]{fontenc}    
\usepackage{url}            
\usepackage{booktabs}       
\usepackage{amsfonts}       
\usepackage{nicefrac}       
\usepackage{microtype}      

\usepackage[english]{babel}

\usepackage{amsmath}
\usepackage{amssymb}
\usepackage{amsthm}

\usepackage{graphicx}

\usepackage{enumitem}

\usepackage[parfill]{parskip}
\usepackage{tabularx}

\usepackage{subcaption}

\newcommand{\calB}{\ensuremath{\mathcal{B}}}
\newcommand{\calD}{\ensuremath{\mathcal{D}}}

\newcommand{\calM}{\ensuremath{\mathcal{M}}}

\newcommand{\calS}{\ensuremath{\mathcal{S}}}
\newcommand{\calU}{\ensuremath{\mathcal{U}}}

\theoremstyle{definition}

\begin{document}

\title{BatPay: a gas efficient protocol for the recurrent micropayment of ERC20 tokens}

\author{Hartwig Mayer \\
	CoinFabrik 
	\And Ismael Bejarano \\
	CoinFabrik
	\And Daniel Fernandez \\
	Wibson
	\And 	Gustavo Ajzenman \\
	Wibson 
	\And Nicolas Ayala \\
	Wibson
	\And Nahuel Santoalla \\
	Wibson
	\And Carlos Sarraute \\
	Wibson \& Grandata
	\And Ariel Futoransky \\
	Disarmista
}

\maketitle

\begin{abstract}

BatPay is a proxy scaling solution for the transfer of ERC20 tokens. It is suitable for micropayments in one-to-many and few-to-many scenarios, including digital markets and the distribution of rewards and dividends.
In BatPay, many similar operations are bundled together into a single transaction in order to optimize gas consumption on the Ethereum blockchain. In addition, some costly verifications are replaced by a challenge game, pushing most of the computing cost off-chain. 
This results in a gas reduction of the transfer costs of three orders of magnitude, achieving around 1700 transactions per second on the Ethereum blockchain.
Furthermore, it includes many relevant features, like meta-transactions for end-user operation without ether, and key-locked payments for atomic exchange of digital goods.

\end{abstract}


\section{Introduction}

In this paper, we present a scaling solution for the transfer of ERC20 tokens called BatPay (short for ``BatchPayment''). 
In this protocol, many similar operations are bundled together into a single transaction in order to optimize gas consumption on the Ethereum blockchain. 
This solution is suitable for micropayments in one-to-many and few-to-many scenarios, including digital markets such as the Wibson data marketplace~\cite{travizano2018wibson,fernandez2020wibson}.
There are three moments when operations are bundled by BatPay:
\begin{itemize}
\item The first is when a Buyer registers a new payment for multiple Sellers in one single transaction.
\item The second is when a Seller collects many payments and sends the tokens to his wallet.
\item The third is when users register in the BatPay platform.
\end{itemize}

\medskip
The main features of BatPay are the following:
\begin{itemize}
\item Cost of 300-1000 gas per payment (depending on operating parameters).
\item No data availability issues, since all the information needed by the players participating in the platform is published onchain.
\item No bottlenecks for normal operation or challenge games.
\item Meta-transactions for ether-less operations of end-users (e.g., the Data Sellers in the Wibson marketplace).
\item Support for immediate withdrawal.
\item Key-locked payments for supporting atomic exchange of digital goods (this feature is optional for Buyers in the platform).
\item Bulk-registration for inexpensive reservation of IDs for new users.
\end{itemize}

The code of the BatPay smart contract is open source and available at: \\
https://github.com/wibsonorg/BatchPayments
 
\section{Related Work}

\paragraph{Payment Pools.} The idea of payment pools was proposed in several places, e.g. \cite{payment_pool_1,payment_pool_2,payment_pool_3}. 
In this approach information about payouts to payees are stored in the leaves of a Merkle
tree. Payees receive via offchain communication the Merkle branch in order to withdraw their
payouts using the correct Merkle proofs. Recurrent payments require an update of the Merkle
tree. To prevent fraudulent updates by the contract owner, a challenge period could be used,
but this would require availability of the necessary offchain data to check correctness.

\paragraph{BatLog.} This concept was proposed in \cite{batog}. BatLog is an efficient mechanism to reward a pool
of participants reflecting their proportion of stake. Instead of iteratively pushing payments, the
total reward is stored in a contract, and users can withdraw their accumulated reward in a pull-based
fashion. The complexity of the algorithm does not depend on the number of payees.
This is a solution for periodic reward distributions, but does not address the general one-to-many
payment problem.

\paragraph{Payment Channels.} In payment channels, as e.g. Raiden, Perun, Counterfactual, Machimony,
Celer, Connext, parties lock a deposit in a contract and send messages containing balance updates via off-chain
channels. When closing the channel, the most recent state update is sent to the blockchain
in a single transaction. During a small window of time, the final state can be challenged. This approach
is only cost-effective for many recurrent uses of the channel, and is preferably applicable for
one-to-few payments. Participants have to be online during the challenge period.

\paragraph{Plasma Chain (Debit / Cash).} The plasma contract acts as a mediator between the root chain and
the Plasma child chain~\cite{plasma_chain}. This contract contains the minimum information so that a user
recognizing fraudulent operations can exit the Plasma chain. Similar to payment channels,
this approach is only suitable for many recurrent payments. Plasma chain assumes availability
of the Plasma chain operator and the chain is vulnerable to mass exits.

\paragraph{Batch Payments using zk-snarks.} In this framework, payers and payees register their addresses and
balance accounts in two different Merkle trees which are stored in a contract~\cite{buterin18}. Users receive
the Merkle branch and can deposit or withdraw from their accounts by providing the correct
Merkle proof. To send a transaction, a user broadcasts an operation. Relayers gather many
of these operations and produce a zk-snark proof that transactions are valid and that the
update of the two Merkle roots is correct. Since nowadays zk-s[nt]arks are very expensive,
the amortization of this approach requires many transactions. In the zk-stark case, proof
construction is often very time-consuming, therefore delaying proof insertion.

Finally, we note that Layer 2 scaling solutions using zk-snarks is an active and promising research field, with projects such as Matter Labs\footnote{\url{https://matter-labs.io/}} among others.

\section{General Overview}

\subsection{BatPay Roles Overview}
\label{sec:roles-overview}

There are five different roles on the BatPay ecosystem: Buyer, Seller, Delegate, Monitor and Unlocker. We provide here an overview and a detailed description in Section~\ref{sec:roles}.
\medskip

\begin{description}
\item[Buyer]
The Buyer deposits tokens into the BatPay contract in order to begin acting as such. She then uses her balance to pay Sellers by registering payments.

\item[Seller]
The Seller participates in several operations with one or many Buyers, and collects afterwards her earnings in her Ethereum account. 

\item[Delegate]
The Delegate simplifies the experience of Sellers by interacting with the BatPay contract on their behalf in exchange for a fee. The Delegate allows Sellers to participate in the protocol without needing gas.
Anyone can operate as Delegate.

\item[Monitor]
The Monitor subscribes to events associated with the BatPay contract, recording outstanding not-yet-collected balances and issuing challenges whenever a Seller or Delegate performs an inconsistent operation.
Anyone can act as Monitor. The Monitor gains a stake when challenging incorrect operations.

\item[Unlocker]
The Unlocker acts as trusted third-party between Seller and Buyer. The Unlocker looks for a locked payment issued by a Buyer, which references a key that she possess, verifies the payment information and, in the case the payment is correct, provides the required key in exchange for a fee.

\end{description}

\subsection{Overview of the BatPay protocol}
\label{sec:roles}

\begin{enumerate}
\item All parties involved in the protocol must register. To identify participants, 32-bit account IDs are used. 
\item The Buyers initiate payments by issuing a \textbf{\emph{registerPayment}} transaction, which includes a per-destination amount and a somewhat-compressed list of Seller IDs. In this transaction, the Buyer indicates the amount that will be paid to each Seller ID (all Sellers receive the same base amount, or an integer multiple of the base amount). The compression consists in specifying the first ID and then only the difference between consecutive IDs (instead of the full ID). 
The total amount paid is deduced from the Buyer's balance.
The Buyer has the option to lock the payment, to ensure that he has access to the intended off-chain asset before the payment occurs.

\item The Sellers wait to accumulate enough payments, then send a \textbf{\emph{collect}} transaction specifying a range of payments and a total amount corresponding to their account. The total amount should be the sum of the payments that the Seller is collecting. 

The collect transaction can be sent through a Delegate. In this way, the Seller can send the collect transaction without using ether.
The Delegate sends a transaction on behalf of the Seller to the BatPay contract specifying which payments have to be transferred to the Seller’s account. The Delegate provides his service in exchange for a fee in tokens.

Data Sellers are encouraged to accumulate payments because of the fee in tokens charged by the Delegate. The more payments he collects in one operation, fewer tokens are charged per payment because the gas consumption of the collect operation remains independent of the number of payments.

\item After a challenge period, the requested amount is added to the Seller's balance.

\item In the case of a dispute (e.g. initiated by a Monitor of the protocol against a particular collect transaction of a particular Seller), the Seller lists the individual payments in which she is included.  The challenger selects a single payment and requests a proof of inclusion. The loser pays for the verification game (stake).
Note that anyone participating in BatPay can act as Monitor.

\end{enumerate}

\section{Detailed Description}

The BatPay contract can be instantiated to act as an optimization proxy for a standard pre-existing ERC20 token contract. User and contract accounts can use it to relay batch micropayments with a hundred times lower gas footprint.

\subsection{Roles Explained}
\label{sec:roles}

\begin{figure}[h]
\begin{center}
{\includegraphics[width=0.85\textwidth]{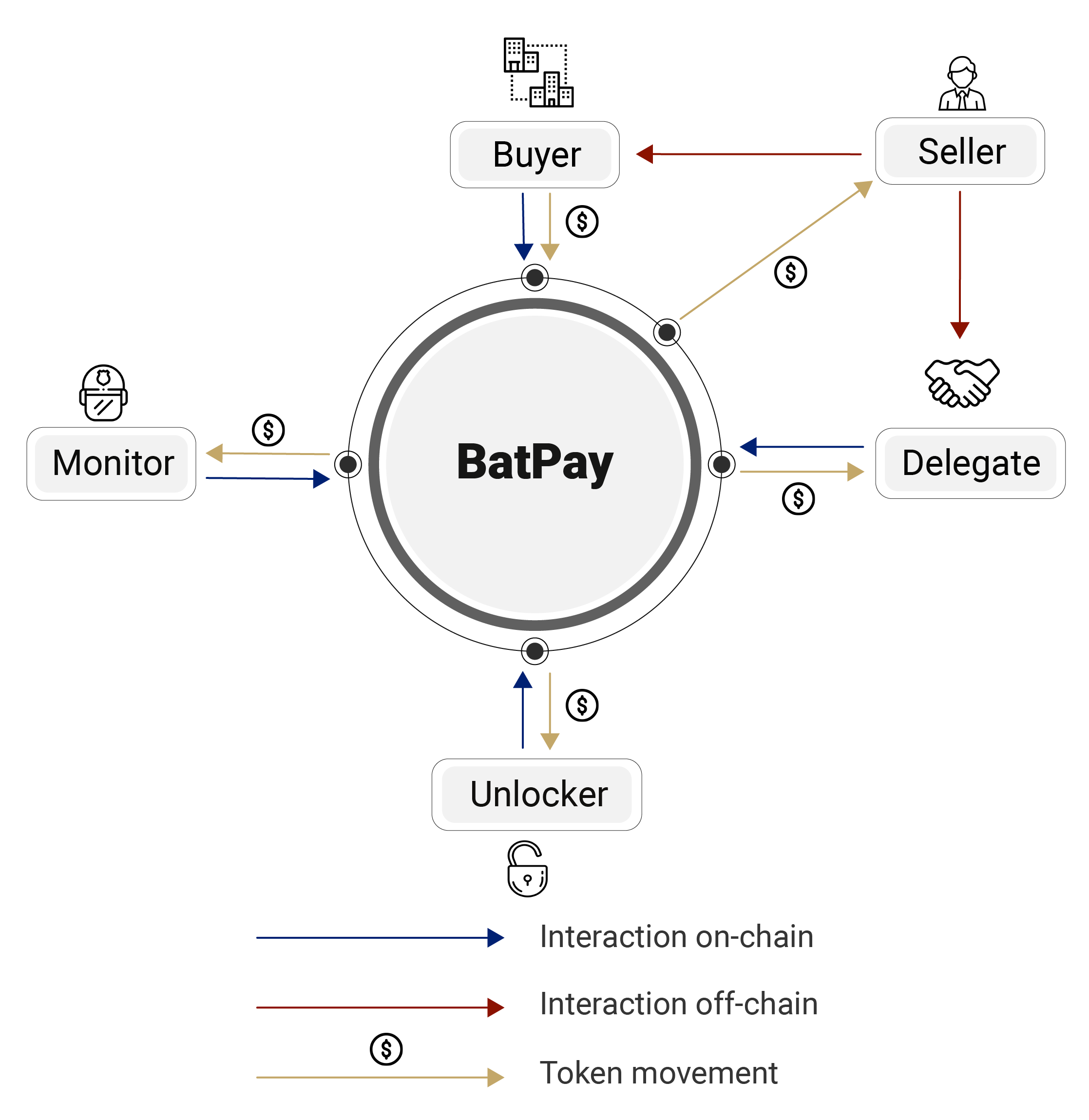} 
}
\caption{Relationships between Buyer, Seller, Unlocker, Delegate and Monitor on the BatPay ecosystem.}
\label{fig:roles}
\end{center}
\end{figure}

We provide here a detailed description of the five different roles on the BatPay ecosystem: Buyer, Seller, Unlocker, Delegate and Monitor (shown in Fig.~\ref{fig:roles}). Players may interact with BatPay in more than one of these roles, depending on the circumstance. All roles will be identified by an account ID obtained during registration.

\begin{description}
\item[Buyer]
The Buyer $\calB$ must deposit tokens into the BatPay contract in order to begin acting as such. She then uses her balance to pay Sellers by issuing registerPayment transactions.

In order to participate in BatPay, a Buyer is required to have:
\begin{itemize}
\item Ethereum address to send and receive payments.
\item Public/private keys for signing transactions.
\item The Buyer must be registered in the BatPay smart contract. Upon registration, he/she receives an identifier $\calB_{ID}$.
\end{itemize}

\item[Seller]
The Seller $\calS$ can participate in several operations with one or many Buyers, and collects afterwards her earnings in her account (with a collect transaction). The balance can be left in BatPay or withdrawn into an ERC20 token account.

In order to participate in BatPay, a Seller is required to have:
\begin{itemize}
\item Ethereum address to send and receive payments~\cite{buterin2014ethereum,wood2014ethereum}. 
\item Public/private keys for signing transactions. 
\item The Seller must be registered in the BatPay smart contract. Upon registration, he/she receives an identifier $\calS_{ID}$.
\end{itemize}

\item[Delegate]
The Delegate $\calD$ simplifies the experience of Sellers by allowing them to interact with the BatPay contract on their behalf in exchange for a fee. The Delegates allow sellers to participate in the protocol without needing gas.
Anyone can operate as delegate.

In order to participate in BatPay, a Delegate is required to have:
\begin{itemize}
\item Ethereum address to send and receive payments.
\item Register in the BatPay smart contract. Upon registration, he/she receives an identifier $\calD_{ID}$.
\item The Delegate must have capital in order to operate in the BatPay system.
\end{itemize}

\item[Monitor]
The Monitor $\calM$ subscribes to events associated with the BatPay contract, recording outstanding not-yet-collected balances and issuing challenges whenever a delegate performs an inconsistent operation.
Anyone can act as monitor. The monitor gains a stake when challenging incorrect operations.

In order to participate in BatPay, a Monitor is required to have:
\begin{itemize}
\item Ethereum address to send and receive payments.
\item Register in the BatPay smart contract. Upon registration, he/she receives an identifier $\calM_{ID}$.
\item The Monitor must have capital in order to operate in the BatPay system.
\end{itemize}

\item[Unlocker]
The Unlocker $\calU$ acts as trusted third-party between Seller and Buyer. 

In BatPay, payments can be locked with a key generated by an Unlocker. This key opens an off-chain asset (that the Buyer received encrypted) at the same time that it releases the payment.
This mechanism is an implementation of the Secure Exchange of Digital Goods~\cite{futoransky2019secure}.

The Unlocker monitors payments, and looks for a locked payment issued by a Buyer, which references a key she possesses.

In that case, the Unlocker verifies the payment information.
The Unlocker performs a set of validations before releasing the payment: that the fee is correct, that the amount of payments is correct, and that the list of payees is correct.

In the case the payment is correct, the Unlocker provides the required key in exchange for a fee. This releases the payment to both Sellers and Unlocker. The key opens an off-chain asset, making it available to the Buyer. 

The Unlocker role is performed by the Notary in the Wibson protocol~\cite{fernandez2020wibson}.

In order to participate in BatPay, an Unlocker is required to have:
\begin{itemize}
\item Ethereum address to send and receive payments.
\item Public/private keys for signing transactions.
\item The Unlocker must be registered in the BatPay smart contract. Upon registration, he/she receives an identifier $\calU_{ID}$.
\end{itemize}

\end{description}

\subsection{Data Structures}

There are four data structures maintained on the smart contract storage: accounts, payments, bulkRecords and collectSlots.

\begin{description}
\item[Account] is an array which stores the Ethereum address, the balance and the latest collected payment associated with an account ID.

\item[BulkRegistration] is an array used to store information about IDs reservation, including the Merkle-tree root hash which will allow a user to later claim an individual ID, and associate it with her address.

\item[Payment] is an array which stores the per-destination amount, the hash of the destination list, as well as other miscellaneous elements associated with each individual payment.

\item[CollectSlot] is a map used to open and manage the collect-challenge game, it stores several attributes associated with the challenge state.

\end{description}

\subsection{Real-world Example}

The version of the Wibson protocol using BatPay improves greatly the efficiency in which it registers or collects payments. We illustrate it with an example. 

Suppose that a Data Buyer registers a payment for 1000 Data Sellers. This transaction consumes 228,255 gas (229 gas per payment). Data Sellers wait to be included in 1000 payments before issuing a collect operation. Collecting 1000 payments consumes 167,440 gas (168 gas per payment). The total amortized cost is 397 gas per individual payment including both transactions. 
Sending the transaction to the network with a Gas Price of 5 Gwei and having an Eth Price of \$225 gives a total cost of \$0.00044 USD.

Our solution improves efficiency by three orders of magnitude, since the gas consumption in the version without BatPay was 400,000 gas per payment, yielding a total of \$0.45 USD for each piece of data exchanged on the market.

\section{Contract Mechanics}

\subsection{Contract Instantiation}

When the BatPay contract is instantiated, the address of a contract implementing the ERC20 interface is supplied. This address is stored to be used later, and cannot be changed. All deposit and withdrawal operations are performed through this address.

\subsection{Users Registration}

Everyone who wants to interact with the BatPay contract, needs to register his address to obtain an associated account ID. This can be performed using one of the available registration methods.

In some cases, paying for registration costs upfront could be prohibitive. For example, the seller may disengage and not participate in a significant number of operations to amortize the registration costs. In this case, direct registration is not attractive and bulkRegistration can be used instead. Bulk Registration can simultaneously register thousands of accounts, while paying for a single transaction cost.

\begin{description}

\item[Direct registration.]
 The registration function can be used to obtain a new account ID, initializing its balance to 0.

\item[Registration on initial deposit.]
Alternatively, executing a deposit operation with a user ID of -1, would register a new account and associate it with the sender address, initializing its balance to the provided amount.

\item[Bulk Registration.]
The \emph{bulkRegistration} function can be used to reserve a range of IDs simultaneously. The sender specifies the number of accounts to reserve and provides the root hash of the Merkle tree holding the list of addresses. This information is saved on contract storage to allow verification.
 
The BulkRegistration is normally performed by a Delegate. The Delegate has incentives to add users to the platform, since it gives him more opportunities to perform collect transactions (and receive its corresponding fee).

At a later time, the \emph{claimBulkRegistrationId} function can be used to assign an address to a pre-reserved account. The sender specifies the account-id, the bulkRegistration-id, and a Merkle proof referencing the address.

\end{description}

\subsection{Payments Mechanism}

Buyers initiate payments by issuing a \textbf{registerPayment} transaction, which includes a per-destination amount and a list of Seller-IDs. In this transaction, the Buyer indicates the amount that will be paid to each Seller-ID (all sellers receive the same base amount, or an integer multiple of the base amount).

The list of Seller-IDs is compressed, it consists in specifying the first ID and then only the difference between consecutive IDs (instead of the full ID). 

The total amount paid is deduced from the Buyer's balance.
The Buyer has the option to lock the payment, to ensure that he has access to the intended off-chain asset before the payment occurs. 

The BatPay contract has a parameter called \emph{unlock period}.
\begin{itemize}
\item If the payment is not locked, it is available for collect after the \emph{unlock period} is elapsed.
\item Whereas if the payment is locked, the Unlocker must present the unlock during the \emph{unlock period}. If this happens, the payment becomes available for collect after the \emph{unlock period} is elapsed. In case of timeout (the unlock is not presented during the specified period), the Buyer gets a refund of his payment.

\end{itemize}

\subsection{Collect Workflow}

The Collect transaction is the moment when the Seller actually gets the tokens corresponding to the payments that she has received in BatPay. 
This is the most complex transaction in the BatPay workflow, 
and involves a Seller $\calS$, a Delegate $\calD$ and a Monitor $\calM$.

\subsubsection{Considerations for the Workflow}
These are the considerations that guided the design of the collect workflow:

\begin{itemize}
\item Sellers are recipient of several small-denomination payment operations (transfer). 

\item Verification for each individual payment could be too expensive to be completed onchain. 

\item Sellers may not have enough funds available to participate on verification games.

\end{itemize}

\subsubsection{Strategy}

A Delegate $\calD$ will get information from a Seller $\calS$ and represent her in a collect/challenge game. The chain-history can be inspected to verify the correct balance for the Seller. The Delegate must be authorized by the Seller to represent him in the collect operation.

The Delegate $\calD$ will issue a collect operation, specifying Seller $\calS$ and its associated balance, and putting a stake for the collect game (the stake is a parameter of the BatPay contract).
The Delegate has the option of performing an Instant Collect, wherein he puts the funds for the Seller, who receives them instantly.
 
Monitors will verify the Seller's balance information and create a challenge if incorrect.

If everything is correct, a transfer will be completed for the Seller.

Collects are verified in individual games (therefore no bottlenecks are generated). There is no data availability problem, since everything is published onchain. 

Note that all players can listen to payments occurring in BatPay, and do their own bookkeeping for all the account IDs. The system is deterministic and both Delegates and Monitors can compute the right balances for all accounts.

\subsubsection{Game State Machine}

\begin{figure}[ht]
\begin{center}
{\includegraphics[width=\textwidth]{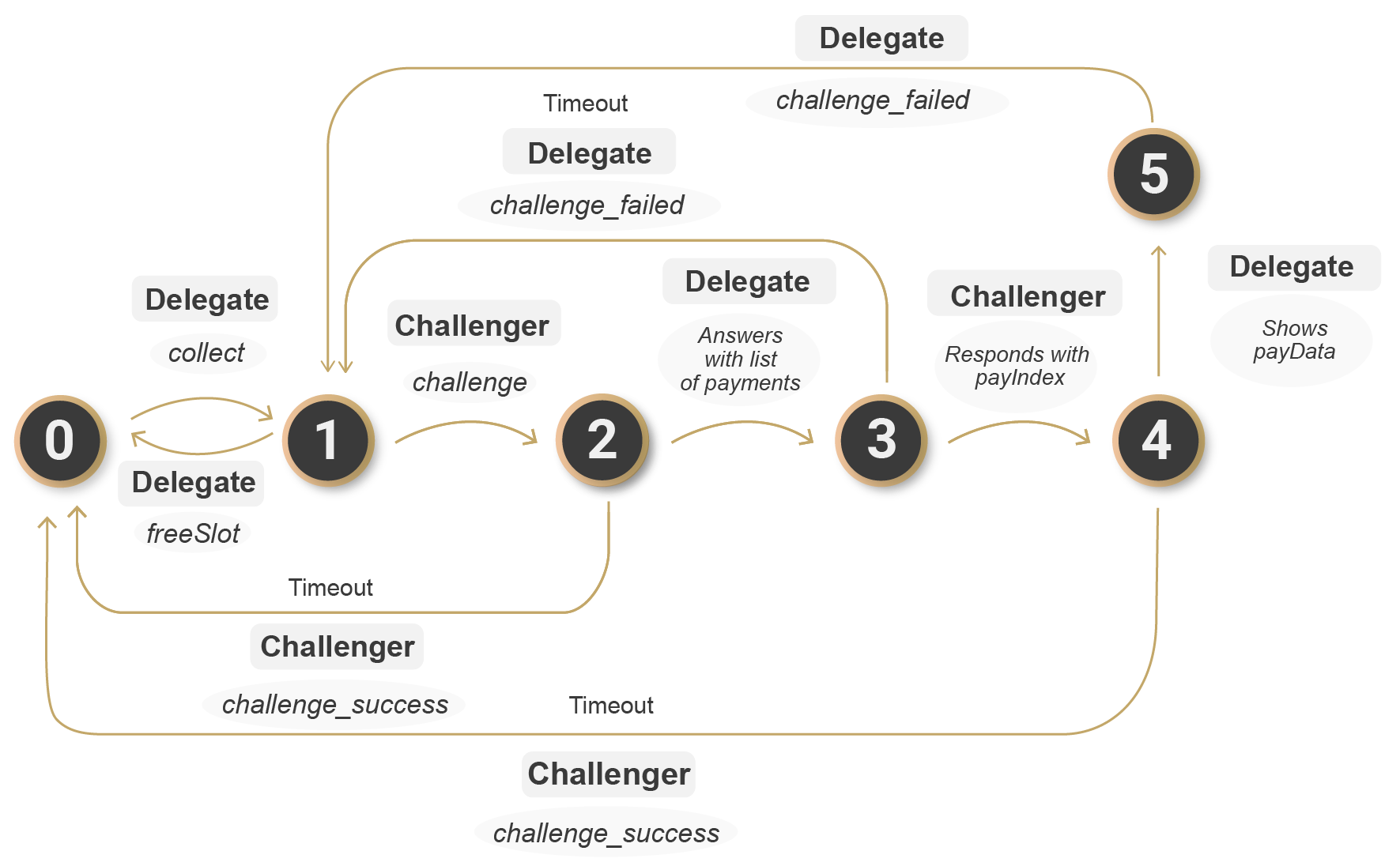} 
}
\caption{Game state machine for the Collect workflow.}
\label{fig:game-state-machine}
\end{center}
\end{figure}

The Delegate $\calD$ must deposit a stake to enter the collect game. The amount of the stake is a parameter of BatPay called \emph{collectStake}. 
The challenger (also called Monitor $\calM$) deposits a stake to enter the collectGame -- the amount is a parameter of BatPay called \emph{challengeStake}.
We describe the states of the collect game state machine (shown in Fig.~\ref{fig:game-state-machine}).

\begin{description}

\medskip
\item[State 0.] Empty slot.

The Delegate $\calD$ calls \emph{collect(delegate, slotId, recipient, lastPaymentIndex, amount, fee, destinationAddress, signature)}, where:

\begin{itemize}
\item \emph{delegate} is the Delegate's ID.
\item \emph{slotId} is the slot's ID to be used during the challenge game.
\item \emph{recipient} is the Seller-ID of the recipient S of the payments.
\item \emph{lastPaymentIndex} is the index of the last payment to be collected
\item \emph{amount} is the sum of the payments to be collected.
\item \emph{fee} is the amount of tokens that the Delegate is charging the Seller to do the transaction. It can be zero.
\item \emph{destinationAddress} (optional) is the wallet where the collected tokens will be transferred. If the address is zero, the tokens will be kept in the Seller's BatPay account.
\item \emph{signature} is the concatenation of the BatPay contract address and all the other parameters in order of appearance, signed by the address stored in the Seller's BatPay account.
\end{itemize}

If the \emph{slotId} is greater than 32,768, then the Seller $\calS$ receives the tokens immediately, but the Delegate $\calD$ must place the amount in return, apart from the \emph{collectStake}. The information of the account contains the index of the last payment already collected (\emph{start = accounts[recipient].collected}).

The collect operation implies collecting all payments for the recipient $\calS$ between the start index and the end index (equal to \emph{lastPaymentIndex}). 

After the call to collect, the game moves to State 1.
 
\medskip
\item[State 1.] Collect game (amount published). 

In this state, the game is waiting for a challenger during a bounded challenge period of time (defined in the BatPay contract).

In case of timeout, the collect operation is successful, and the Seller $\calS$ receives the tokens. 
In that case, the Delegate $\calD$ may call \emph{freeSlot()}, thus returning to State 0, and liberating the slot to start another collect game. 

For a challenge to happen, a Monitor $\calM$ must call \emph{challenge\_1(delegateId, slotId, challengerId)}, which moves the game to State 2.

\medskip
\item[State 2.] Challenge started. 

After the challenge starts, the Delegate $\calD$ has a bounded time period to respond to the challenge (specified in the BatPay contract).

In case of timeout or if the Delegate fails to respond, the challenge is successful for the Monitor $\calM$. In that case, by calling \emph{challenge\_success(delegateId, slotId)}, the Delegate loses his stake, the recipient $\calS$ does not receive the funds, and the Monitor $\calM$ earns tokens for his successful challenge. The game returns to State 0.

If the Delegate responds with a list of pay indexes and amounts corresponding to the recipient: 
$$ 
\left[ \; (\mbox{\emph{payIndex}}_0, \mbox{\emph{amount}}_0 ), (\mbox{\emph{payIndex}}_1, \mbox{\emph{amount}}_1 ), \ldots, (\mbox{\emph{payIndex}}_n, \mbox{\emph{amount}}_n ) \; \right]
$$
then the game moves to State 3. 

\medskip
\item[State 3.] Waiting for individual payment selection. 

The challenger (i.e. the Monitor $\calM$) has a bounded period of time to single out a payment that he considers incorrect.

In case of timeout, the challenge failed, and the machine returns to State 1, by calling \emph{challenge\_failed(delegateId, slotId)}.
 
If the challenger $\calM$ gives the information $(\mbox{\emph{payIndex}}_k, \mbox{\emph{amount}}_k )$ of the payment that he is challenging, then the game moves to State 4. 

\medskip
\item[State 4.] Waiting for proof for $\mbox{\emph{payIndex}}_k $.

The Delegate $\calD$ has a bounded period of time to respond with the \emph{payData} corresponding to $\mbox{\emph{payIndex}}_k $.

In case of timeout or if the Delegate fails to respond, the challenge is successful for $\calM$, and the game returns to State 0, again by calling the \emph{challenge\_success} method.

If the Delegate gives the \emph{payData} for $ \mbox{\emph{payIndex}}_k $ with the right amount, thereby demonstrating that the payment is correct, then the game moves to State 5.

\medskip
\item[State 5.] Proof for $\mbox{\emph{payIndex}}_k $ is correct.

In the case the proof is correct, the challenge of $\calM$ failed and the game returns to State 1, by calling the \emph{challenge\_failed} method.

\end{description}

\section{Conclusion}

The Wibson data marketplace presented in \cite{travizano2018wibson,fernandez2020wibson} will benefit consumers by providing them the ability to control and monetize their personal information. It will also give access to high quality and verified data to organizations which need to train Machine Learning algorithms and models, as well as an explicit consumer consent mechanism which will be absolutely critical as new privacy regulations are coming into effect.

In addition to the marketplace protocol, Wibson also provides primitives to solve efficiently the problem of fair exchange~\cite{cleve1986limits} by providing an efficient mechanism for the secure exchange of digital goods~\cite{futoransky2019secure}, reminiscent of the \emph{secure triggers} cryptographic primitive~\cite{futoransky2006foundations}.
 
In this paper we presented an improvement to the Wibson protocol called BatPay, wherein many similar operations are bundled together into a single transaction in order to optimize gas consumption on the Ethereum blockchain. In addition, some costly verifications are replaced by a challenge game, pushing most of the computing cost off-chain. 

This results in a gas reduction in transfer costs of about three orders of magnitude. 
Furthermore, BatPay includes many relevant features, like meta-transactions for end-user operation without ether, and key-locked payments for the atomic exchange of digital goods~\cite{futoransky2019secure}.

\subsection*{Code availability}

The code of the BatPay smart contract is open source and available at: \\
https://github.com/wibsonorg/BatchPayments

\subsection*{Acknowledgements}

The authors thank Juan Carrillo for his work in the development of the Wibson platform.

{ \small 
\bibliographystyle{unsrt}
\bibliography{../wibson}

\begin{thebibliography}{10}

\bibitem{travizano2018wibson}
Matias Travizano, Carlos Sarraute, Gustavo Ajzenman, and Martin Minnoni.
\newblock Wibson: A decentralized data marketplace.
\newblock In {\em Proceedings of SIGBPS 2018 Workshop on Blockchain and Smart
  Contract}, 2018.

\bibitem{fernandez2020wibson}
Daniel Fernandez, Ariel Futoransky, Gustavo Ajzenman, Matias Travizano, and
  Carlos Sarraute.
\newblock Wibson protocol for secure data exchange and batch payments.
\newblock {\em arXiv 2001.08832}, 2020.

\bibitem{payment_pool_1}
Nick Johnson.
\newblock How to send ether to 11,440 people.
\newblock {\footnotesize
  \url{https://medium.com/@weka/how-to-send-ether-to-11-440-people-187e332566b7}},
  2016 (accessed Feb 4, 2020).

\bibitem{payment_pool_2}
Peter Watts.
\newblock Pooled payments.
\newblock {\footnotesize
  \url{https://ethresear.ch/t/pooled-payments-scaling-solution-for-one-to-many-transactions/590}},
  2018 (accessed Feb 4, 2020).

\bibitem{payment_pool_3}
Hassan Abdel-Rahman.
\newblock Scalable payment pools in solidity.
\newblock {\footnotesize
  \url{https://medium.com/cardstack/scalable-payment-pools-in-solidity-d97e45fc7c5c}},
  2018 (accessed Feb 4, 2020).

\bibitem{batog}
Bogdan Batog, Lucian Boca, and Nick Johnson.
\newblock Scalable reward distribution on the ethereum blockchain.
\newblock {\footnotesize
  \url{http://batog.info/papers/scalable-reward-distribution.pdf}}, 2018
  (accessed Feb 4, 2020).

\bibitem{plasma_chain}
Anton Bukov.
\newblock Eip1035: Transaction execution batching and delegation.
\newblock {\footnotesize \url{https://github.com/ethereum/EIPs/issues/1035}},
  2018 (accessed Feb 4, 2020).

\bibitem{buterin18}
Vitalik Buterin.
\newblock On-chain scaling to potentially ~500 tx/sec through mass tx
  validation.
\newblock {\footnotesize
  \url{https://ethresear.ch/t/on-chain-scaling-to-potentially-500-tx-sec-through-mass-tx-validation/3477}},
  2018 (accessed Feb 4, 2020).

\bibitem{buterin2014ethereum}
Vitalik Buterin.
\newblock Ethereum: A next-generation smart contract and decentralized
  application platform.
\newblock {\em Ethereum White Paper}, 2014.

\bibitem{wood2014ethereum}
Gavin Wood.
\newblock Ethereum: A secure decentralised generalised transaction ledger.
\newblock {\em Ethereum Project Yellow Paper}, 151, 2014.

\bibitem{futoransky2019secure}
Ariel Futoransky, Carlos Sarraute, Ariel Waissbein, Daniel Fernandez, Matias
  Travizano, and Martin Minnoni.
\newblock Secure exchange of digital goods in a decentralized data marketplace.
\newblock In {\em Proceedings of the 2019 Argentine Symposium on Big Data
  (AGRANDA)}, pages 38--44, 2019.

\bibitem{cleve1986limits}
Richard Cleve.
\newblock Limits on the security of coin flips when half the processors are
  faulty.
\newblock In {\em Proceedings of the eighteenth annual ACM symposium on Theory
  of computing}, pages 364--369. ACM, 1986.

\bibitem{futoransky2006foundations}
Ariel Futoransky, Emiliano Kargieman, Carlos Sarraute, and Ariel Waissbein.
\newblock Foundations and applications for secure triggers.
\newblock {\em ACM Transactions on Information and System Security (TISSEC)},
  9(1):94--112, 2006.

\end{thebibliography}
}

\end{document}